\documentclass[oneside,11pt]{amsart}

\usepackage{amsmath,amssymb,epsf}
\usepackage{amsfonts,amsbsy,amscd,stmaryrd}
\usepackage{latexsym,amsopn}
\usepackage{amsthm}
\usepackage{esint}
\usepackage{mathrsfs}
\allowdisplaybreaks
 \usepackage{enumitem}
\usepackage{color}
\usepackage{cancel}

\newcounter{smallarabics}
\newenvironment{arabicenumerate}
{\begin{list}{{\normalfont\textrm{(\arabic{smallarabics})}}}
  {\usecounter{smallarabics}\setlength{\itemindent}{0cm}
   \setlength{\leftmargin}{5ex}\setlength{\labelwidth}{4ex}
   \setlength{\topsep}{0.75\parsep}\setlength{\partopsep}{0ex}
   \setlength{\itemsep}{0ex}}}
{\end{list}}

\newcounter{smallroman}

\newcommand{\ben}{\begin{arabicenumerate}}  
\newcommand{\een}{\end{arabicenumerate}}

\newtheorem{theorem}{Theorem}[section]
\newtheorem{assumption}{Hypothesis}[section]
\newtheorem{proposition}[theorem]{Proposition}
\newtheorem{lemma}[theorem]{Lemma}
\theoremstyle{definition}
\newtheorem{definition}[theorem]{Definition}
\newtheorem{corollary}[theorem]{Corollary}
\newtheorem{remark}[theorem]{Remark}
\newtheorem{example}[theorem]{Example}
\newcommand{\beq}{\begin{equation}}
\newcommand{\eeq}{\end{equation}}
\newcommand{\beqa}{\begin{eqnarray}}
	\newcommand{\eeqa}{\end{eqnarray}}
\newcommand{\bea}{\begin{aligned}}
\newcommand{\eea}{\end{aligned}}

\newcommand{\bex}{\begin{example}}
\newcommand{\eex}{\end{example}}
\def\bel{\begin{lemma}}
\def\eel{\end{lemma}}
\def\bet{\begin{theorem}}
\def\eet{\end{theorem}}
\def\bed{\begin{definition}}
\def\eed{\end{definition}}
\def\ber{\begin{remark}}
\def\eer{\end{remark}}

\def\rr{{\mathbb R}}

\def\ss{{\mathbb S}}

\def\part{{\rm par}}

\def\cpl{{\rm cpl}}

\def\proof{
\noindent{\bf Proof.}\ \ }

\usepackage{calrsfs}
\DeclareMathAlphabet{\pazocal}{OMS}{zplm}{m}{n}

\def\cY{{\pazocal Y}}
\def\cZ{{\pazocal Z}}

\def\cD{{\pazocal D}}

\def\cC{{\pazocal C}}

\let\dotlessi\i
\def\i{{\rm i}}

\def\loc{{\rm loc}}

\DeclareMathOperator{\Ker}{Ker}

\renewcommand\ker{{\rm ker}}

\def \p{ \partial}

\def\12{\frac{1}{2}}
\def\14{\frac{1}{4}}

\def\e{{\rm e}}

\DeclareMathOperator{\Ran}{Ran}
\DeclareMathOperator{\supp}{supp}

\newcommand{\one}{\boldsymbol{1}}

\def\c{{\rm c}}

\def\cC{{\pazocal C}}

\def\cX{{\pazocal X}}

\def\12{\frac{1}{2}}

\newcommand{\mfh}{\mathfrak{h}}

\def\e{{\rm e}}

\def\Ran{{\rm Ran}}

\def\Diff{{\rm Diff}}
\def\bx{{\rm x}}

\def\bep{\begin{proposition}}
\def\eep{\end{proposition}}
\def\b{{\rm b}}

\newcommand{\Ga}{\Gamma}

\newcommand{\la}{\lambda}

\DeclareSymbolFont{boldoperators}{OT1}{cmr}{bx}{n}
\SetSymbolFont{boldoperators}{bold}{OT1}{cmr}{bx}{n}

\makeatletter
\newcommand*{\defeq}{\mathrel{\rlap{%
                     \raisebox{0.3ex}{$\m@th\cdot$}}%
                     \raisebox{-0.3ex}{$\m@th\cdot$}}%
                     =}
                     
\makeatother
\makeatletter
\newcommand*{\eqdef}{=\mathrel{\rlap{%
                     \raisebox{0.3ex}{$\m@th\cdot$}}%
                     \raisebox{-0.3ex}{$\m@th\cdot$}}%
                     }
\makeatother
\newcommand{\traa}[1]{\mskip-6mu\upharpoonright_{#1}}

\def\cf{\pazocal{C}^\infty}

\def\pM{{\p M}}

\def\c{{\rm c}}

\def\inti{{\circ}}

\def\loc{{\rm loc}}
\newcommand{\Hl}{H_{0,\b,\loc}}

\newcommand{\Hc}{H_{0,\b,{\rm c}}}

\newcommand{\Hpm}{H_{0,\b,\pm}}

\let\origmaketitle\maketitle
\def\maketitle{
  \begingroup
  \def\uppercasenonmath##1{} 
  \let\MakeUppercase\relax 
	\origmaketitle
  \endgroup
}

\newcommand{\AdS}{{\rm AdS}}

\newcommand{\dv}{{([0,\epsilon)_z;\cD'(\p M))}}

\def\pX{\partial M}
\def\pM{\partial M}

\def\mass{\nu^2 - \textstyle\frac{(n-1)^2}{4}}

\def\bulk{{\rm bulk}}

\def\bd{{\rm bd}}
\newcommand{\si}{\sigma}
\newcommand{\real}{\mathbb{R}}

\newcommand{\mrm}{\mathrm}

\newcommand{\mfa}{\mathfrak{A}}
\newcommand{\mfA}{\pazocal{A}}

\newcommand{\res}{\restriction}

\newcommand{\nin}{\noindent}

\newcommand{\ti}{\tilde}

\newcommand{\hil}{\pazocal{H}}
\newcommand{\ov}{\overline}
\newcommand{\Om}{\Omega}
\newcommand{\om}{\omega}

\begin{document}

\title[A mechanism for holography on AdS spacetimes]{\large A mechanism for holography for non-interacting fields\\ on anti-de Sitter spacetimes}

\author{\large Wojciech Dybalski \& Micha{\l} {Wrochna}}
\address{Technische Universit\"at M\"unchen, Boltzmannstrasse 3, 85748 Garching, Germany}
\email{dybalski@ma.tum.de}

\address{Universit\'e Grenoble Alpes, Institut Fourier, UMR 5582 CNRS, CS 40700, 38058 Grenoble \textsc{Cedex} 09, France}
\email{michal.wrochna@univ-grenoble-alpes.fr}
\keywords{Quantum Field Theory on curved spacetimes, asymptotically anti-de Sitter spacetimes, holography, Hadamard condition}

\thanks{\emph{Acknowledgments.} The authors are grateful to Maximilian Duell, Ko Sanders, Arick Shao and Aron Wall for useful discussions. Support from the grant ANR-16-CE40-0012-01 and the grant \textsc{FK13\textunderscore\,2017} of the French-Bavarian Center for Academic Cooperation are gratefully acknowledged.}

\begin{abstract} In the setting of non-interacting Klein-Gordon fields on asymptotically anti-de Sitter spacetimes, we show that algebras of observables localized in a neighborhood of the boundary are subalgebras of a boundary algebra. The underlying mechanism is directly related to holography for classical fields. In particular, the proof relies on unique continuation theorems at the conformal boundary. 
\end{abstract}

\maketitle

\section{Introduction}\label{sec:introduction}

\subsection{Introduction, main result} Manifestations of the AdS/CFT duality can be efficiently probed on the level of classical fields. In particular, given a fixed asymptotically anti-de Sitter  (aAdS)  spacetime $(M,g)$, it is well understood how to assign to each classical solution of the Klein-Gordon equation a datum on the conformal boundary $\pM$ \cite{grahamlee,vasy}. Furthermore, it was shown recently by Holzegel \& Shao that solutions with vanishing data necessarily vanish in a bulk neighborhood of the boundary \cite{HS1,HS2}. Alongside with Holmgren's theorem in the analytic case, this provides a rigorous mechanism for holography in the classical case.

Now, in the case of quantum fields, one can ask how holography is tied to the classical degrees of freedom, see for instance \cite{KL}. The question is interesting even for non-interacting fields on a fixed aAdS background, which are believed to approximate the $N\to\infty$ limit of Maldacena's form of AdS/CFT for small field amplitudes \cite{maldacena}. It can also be asked in the rigorous setting of \emph{Rehren duality} \cite{rehren,rehren2} (or \emph{algebraic holography}, cf.~\cite{ribeiro1,ribeiro2} for a generalization), which relates two quantum theories without explicitly appealing to classical dynamics. A clearer link with the classical case is exhibited by what is often called the  \emph{field-theoretical correspondence}, which is however based on a boundary-to-bulk map that integrates over different field masses; its interpretation is thus less direct, cf.~the works of D\"utsch and Rehren, \cite{DR1} and  \cite[(4.17)]{DR2}. Multiple indications about the classical phenomena underpinning quantum holography can be found in the physics literature, see e.g.~\cite{HKLL,morrison,KW,KO}, cf.~\cite{harlow} for a recent review including bulk reconstruction procedures. However, typically high symmetry of spacetime is assumed and the analysis relies on explicit formul\ae, which somewhat obscures the existence of a general mechanism. 

\medskip

In the present work we focus on linear Klein-Gordon quantum fields on a fixed aAdS spacetime. Our goal is to provide a mechanism for quantum holography based directly on a unique continuation property for classical fields of the kind considered by Holzegel \& Shao or as in Holmgren's theorem in the analytic case. Our main result can be summarized as follows.

For any open set $O$ in the boundary $\p M$, let $V(O)\subset M$ be defined by the property that any classical solution with data vanishing on $O$ automatically vanishes on $V(O)$. Then we have the inclusion of $C^*$-algebras
\beq\label{eq:main}
\mfa_{\bd}(O)\supset\mfa_{\bulk}(V(O)), 
\eeq
where $\mfa_{\bd}(O)$ represents boundary observables localized in the region $O$, and $\mfa_{\bulk}(V(O))$ are bulk observables localized in $V(O)$. 

In our non-interacting setting, the algebras \eqref{eq:main} are CCR $C^*$-algebras. They are spanned by abstract Weyl operators $W(v)$, where $v$ are distributions supported in $\pM\times O$, respectively in $V(O)$. The main difficulty is that distributions of the form $v=\delta_{\pM}\otimes \varphi$ with $\varphi\in C_{\rm c}^{\infty}(O)$ are too singular to belong to the symplectic space of functions on $M$ obtained by generalizing the standard constructions to aAdS spacetimes. To cope with this problem we take the completion with respect to the scalar product induced e.g.~by a quasi-free state $\omega$. Whereas the choice of $\omega$ enters the definition of $\mfa_{\bulk}(V(O))$, the result is valid for any state $\omega$ satisfying a mild regularity assumption (analogous to the often made requirement on globally hyperbolic spacetimes that the $n$-point functions are distributions). In particular, it is satisfied by all \emph{holographic Hadamard states} \cite{wrochna}. Therefore, \eqref{eq:main} reflects features of the system that are independent on the choice of particular $\omega$.

We remark that \eqref{eq:main} implies that in any representation $\pi$, if a vector $\Omega$ is cyclic for $\pi(\mfa_{\bulk}(V(O)))$ then it is also cyclic for $\pi(\mfa_{\bd}(O))$. In particular, if $V(O)\neq\emptyset$ and if the Reeh-Schlieder property holds true in the bulk, then a dense subspace of the Hilbert space can be recovered by acting on $\Omega$ with boundary observables, as in the work of Morrison \cite{morrison}.

\subsection{Plan of the paper} To make our proofs more transparent, we first work in Section \ref{sec:abstract} in a completely abstract setup. Theorem \ref{Main-result} therein is the abstract version of our main result, and is expected to apply in a variety of different settings. We also give an alternative to Theorem \ref{Main-result} for $\omega$-independent bulk algebras. In Section \ref{sec:AdS} we specialize to (Dirichlet) Klein-Gordon fields on asymptotically AdS spacetimes and we check that the assumptions of Theorem \ref{Main-result} are satisfied. This part is largely based on the analysis in \cite{wrochna}, which in turn relies on results by Vasy on classical fields \cite{vasy}. In the last part of Section 3 we explain how \eqref{eq:main} applies with non-empty $V(O)$ in the analytic case, and provide an outlook concerning the smooth case, in which the use of Holzegel and Shao's result encounters extra technical difficulties.

Appendix A briefly reviews Weyl CCR $C^*$-algebras and quasi-free representations.

\section{Abstract setup}\label{sec:abstract}

\subsection{Notation} If $\cX$ is a topological real vector space, we denote by $\cX'$ its continuous dual equipped with the weak$^*$ topology. 

If $\cY$ is a topological real vector space and $\alpha:\cX\to\cY$ is a continuous $\mathbb{R}$-linear map, we denote by $\alpha':\cY'\to\cX'$ the dual operator. If $\cZ\subset \cX$ is a subspace, then we denote by $\alpha\restriction \cZ$ the restriction of $\alpha$ to $\cZ$ meant as an operator $\cZ\to \cY$. 

The space of bi-linear forms on $\cX$  is denoted by $L(\cX,\cX')$, and the subspace of symmetric ones by $L_{\mrm{s}}(\cX, \cX')$. For $\eta\in L(\cX,\cX')$, we denote by $( v_1 \cdot \eta v_2)\in\rr$ the evaluation of $\eta$ on $v_1,v_2\in\cX$.

\subsection{Framework and main result}\label{ss:abstract}

We introduce an abstract setting to study relations between the algebras of observables of the free scalar field 
localized at the boundary and at the bulk of a Lorentzian spacetime. Namely:

\begin{enumerate}[label={(\arabic*)}]
\item Let $\cX$ be a  barreled\footnote{See e.g.~\cite[Def.~3.1, 33.1]{treves} or \cite{Ja81} for the definition of a barreled space. For our applications it is sufficient to consider Fr\'echet spaces and inductive limits of Fr\'echet spaces, which are all barreled.} topological real vector space, and let $\si$ be a symplectic form on $\cX$. Let $\cX_\bulk\subset \cX$ be a subspace of $\cX$.
\item Let $\cX_{\bd}$ be a topological real vector space.
\item Let $\p: \cX_{\bulk}'\to \cX_{\bd}'$ be a linear map such that the restriction \\  $\p\res \cX'$ is continuous.
\item Let $\eta\in L_{\mrm{s}}(\cX, \cX')$ be strictly positive, i.e.~$(v\cdot \eta v )>0$ if $v\neq 0$.  Furthermore, we assume that $|v_1\cdot \sigma v_2|\leq (v_1\cdot \eta v_1)^{1/2} (v_2\cdot \eta v_2)^{1/2}$ for all $v_1,v_2\in \cX$.
\end{enumerate}

The space $\cX$ represents \emph{all} bulk degrees of freedom.  We will focus our attention on a possibly smaller set of bulk degrees of freedom $\cX_{\bulk}$. The space $\cX_{\bd}$ describes some boundary degrees of freedom. 

 The positivity conditions on the symmetric form $\eta$ will allow us to interpret it as the covariance of a quasi-free state, see Appendix \ref{theappendix} for a brief introduction on CCR algebras and quasi-free states.

In the first place, we want to embed $\cX_{\bd}$ in bulk degrees of freedom. 
 Some special care is needed, however, because in practice we expect the embedding to make sense only after replacing $\cX$ by its completion $\cX^\cpl$ with respect to $\eta$. Such completion arises naturally in the GNS construction w.r.t.~the quasi-free state associated to $\eta$.

Then, the goal will be to study the relationship between $\cX_\bd$ and $\cX_\bulk$. In what follows we consider a regularity assumption that allows us to construct the embedding (Hypothesis \ref{hyp1}), and a further hypothesis which will turn out to imply an inclusion of associated algebras (Hypothesis \ref{hyp2}).


\begin{assumption}\label{hyp1} The map $\eta: \cX \to \cX'$ is continuous. \end{assumption}

\begin{assumption}\label{hyp2} For all $u\in \cX'$ we have the implication:
\beqa\label{eq:hyp2}
\p u=0 \ \ \Rightarrow \ \ (u=0 \textrm{ as a functional on  } \cX_{\bulk} ).
\eeqa
\end{assumption}

\nin The `injectivity' property \eqref{eq:hyp2} is the most important ingredient for our analysis. In essence, our main result 
in Theorem~\ref{Main-result} reflects the  general fact, that the dual of an injective map  on a 
locally convex Hausdorff  space has a weak$^*$ dense range (cf.~\cite[Sec.~V, Problem 19]{RS1}). However, we cannot apply this 
general property directly, since a priori $\p':  \cX_{\bd}''\to \cX_{\bulk}''$  maps outside of the domain of definition of quantum fields. Another difficulty is the restriction to the subspace $\cX_{\bulk}$. These 
problems are resolved in the next lemma.

We stress that $\cX$ and $\cX^\cpl$ carry in general different topologies; when writing $\cX^\cpl$ we will always mean the topology defined by $\eta^\cpl$ (the scalar product induced by $\eta$). 

\bel\label{technical-lemma} Assume Hypothesis \ref{hyp1}. Then
\begin{equation}\label{eq:tl1} (\Ran\,( \p' \res   \cX_{\bd}))\subset \cX^{\cpl}.
\end{equation}
Furthermore, if Hypothesis \ref{hyp2} is satisfied then
\begin{equation}\label{eq:tl2}
(\Ran\,( \p' \res   \cX_{\bd}))^{\mrm{cl}}\supset (\cX_{\bulk})^{\mrm{cl}}, 
\end{equation}
where the closures are in the Hilbert space topology given by $\eta$. 
\eel
\proof Let us first show that 
\begin{equation}\label{eq:etac}
\eta^{\cpl}: \cX^{\cpl} \to \cX' \ \mbox{continuously}.
\end{equation} 
For all $\ti u\in \cX^{\cpl}$, $v\in \cX$,  we have by the Cauchy-Schwarz inequality and the positivity of $\eta^{\cpl}$ 
\beqa
\begin{aligned}\label{Cauchy-Schwarz}
|(v\cdot \eta^{\cpl} \ti u)|&\leq (v\cdot \eta^{\cpl} v)^{1/2}
(\ti u\cdot \eta^{\cpl}\ti u)^{1/2}\\ &= (v\cdot \eta v)^{1/2}
(\ti u\cdot \eta^{\cpl}\ti u)^{1/2}.
\end{aligned}
\eeqa
If $\cX\ni v_n\to 0$ then $(v_n\cdot \eta v_n)\to 0$ by Hypothesis \ref{hyp1}, and the fact that 
a separately continuous linear functional on a  barreled space is jointly continuous \cite[Thm.~41.2]{treves}. Hence $(v_n\cdot \eta^{\cpl} \ti u)\to 0$ by \eqref{Cauchy-Schwarz}. This proves that $\eta^{\cpl}$ maps to $\cX'$.

Now, suppose that  $\cX^{\cpl}\ni \ti u_n\to 0$, i.e. $(\ti u_n\cdot \eta^{\cpl} \ti u_n)\to 0$. Then, by \eqref{Cauchy-Schwarz}, $\eta^{\cpl} \ti u_n\to 0$ in the weak$^*$-topology. This shows \eqref{eq:etac}. \medskip

\nin \emph{Proof of \eqref{eq:tl1}:} \, Let $f\in \cX_{\bd}$.  We define the linear functional  
\beqa
 \cX^{\cpl}\ni u\mapsto (\p' f\cdot \eta^{\cpl} u)
=(f\cdot \p\eta^{\cpl} u),
\eeqa
which is continuous because $\p\circ\eta^{\cpl}: \cX^{\cpl} \to \cX_{\bd}'$  is continuous by \eqref{eq:etac} and continuity of $\p: \cX'\to\cX_\bd'$. By the Riesz representation theorem this implies $\p'f\in \cX^{\cpl}$. 

\medskip

\nin \emph{Proof of \eqref{eq:tl2}}: \, Suppose $u \in \cX^{\cpl}$ is orthogonal to all $\p'f$, $f\in  \cX_{\bd}$, i.e.~$( u\cdot \eta^{\cpl}  \p'f)=0$. Define a functional $g_u\in \cX'$ by $g_{u}(w)\defeq (u\cdot \eta^{\cpl} w)$ for all $w\in \cX$.
Then $\cX'_{\bd} \ni \p g_{u}=0$ and, by Hypothesis \ref{hyp2}, $g_{u}=0$  on all elements of $\cX_{\bulk}$, that is $(u\cdot \eta^{\cpl} w_{\bulk})=0$
for all $w_{\bulk}\in \cX_{\bulk}$. Thus, we have shown that
\beqa
(\Ran\,( \p' \res   \cX_{\bd}))^{\bot}\subset  (\cX_{\bulk})^{\bot},
\eeqa
which gives the claim. \qed\\

We remark that the argument we use to show \eqref{eq:tl2} works similarly to the proof of the Reeh-Schlieder property for analytic Hadamard states on curved spacetimes due to Strohmaier, Verch and Wollenberg \cite{SVW}.

\medskip

Let $\si^\cpl$ be the pre-symplectic form induced by $\sigma$ on the completion $\cX^\cpl$. To state our main result we define the 
$C^*$-algebra $\mfa\defeq \mrm{CCR} (\cX^{\cpl}, \si^{\cpl})$ (see Appendix A) and the following sub-algebras
\begin{equation}\begin{aligned}
\mfa_{\bulk}&\defeq \mrm{CCR} (\cX^{\mrm{cl}}_{\bulk}, \si^{\cpl}),\\
\mfa_{\bd}&\defeq \mrm{CCR} (\cX_{\bd}^{\p}, \si^{\cpl}), \textrm{ where } \cX_{\bd}^{\p}\defeq  (\Ran\,( \p' \res   \cX_{\bd}))^{\mrm{cl}}.\end{aligned}
\end{equation}
The closures refer to the $\cX^{\cpl}$ topology. We note that by Lemma~\ref{technical-lemma}, the boundary algebra is well defined as $\cX_{\bd}^\p\subset \cX^{\cpl}$. As an immediate consequence of \eqref{eq:tl2} we obtain:
\bet\label{Main-result} Assume Hypotheses \ref{hyp1} and \ref{hyp2}. Then $\mfa_{\bd}\supset \mfa_{\bulk}$. 
\eet


\begin{corollary} As a consequence of Theorem \ref{Main-result}:
\begin{enumerate}
\item[1.] In any representation $\pi$ (not necessarily related to $\eta$) we have 
\[
\pi(\mfa_{\bulk})\subset \pi(\mfa_{\bd}) \mbox{ and } \pi(\mfa_{\bulk})''\subset \pi(\mfa_{\bd})'',
\] 
where the prime denotes here the commutant. 
\item[2.] Any vector which is cyclic for $\pi(\mfa_{\bulk})$ is also cyclic for $\pi(\mfa_{\bd})$. 
\end{enumerate}
\end{corollary}

\begin{remark}\label{theremark} Hypothesis \ref{hyp2} can be weakened at the cost of strengthening Hypothesis \ref{hyp1}. Namely, suppose that $\cY$ is a locally convex topological vector space such that $\cX\subset \cY \subset \cX^\cpl$, and assume Hypotheses \ref{hyp1}-\ref{hyp2} and (3) with $\cX$ replaced by $\cY$ (thus with $\cX'$ replaced by $\cY'$). Then the assertion of Theorem \ref{Main-result} is still true, as can be shown by a straightforward modification of the proof.
\end{remark} 

\subsection{Formulation in Kay-Wald representation}

 Let us now give an alternative algebraic construction, whose advantage is that no information about the  `state' (i.e.~$\eta$)  enters into the bulk algebra.

Consider the $\eta$-independent $C^*$-subalgebras 
\begin{align}
\mfA&\defeq  \mrm{CCR} (\cX, \si)\subset \mfa, \\
\mfA_{\bulk}&\defeq  \mrm{CCR} (\cX_{\bulk}, \si) \subset \mfa_{\bulk}.
\end{align}
Although for $\mfa_{\bd}$ we do not have an $\eta$-independent counterpart, we can  prove a variant of Theorem~\ref{Main-result} 
in the Kay-Wald representation $(\hil_{\mrm{KW}}, \pi_{\mrm{KW}}, \Om_{\mrm{KW}})$ w.r.t.~$\eta$, see Proposition~\ref{GNS-proposition} in the appendix. To that end we have to suppose in addition that $\eta$ defines a quasi-free state, so we need to assume the second property in (\ref{positivity-property}) on $\eta$. 
 Also, to meet the assumptions of Proposition~\ref{GNS-proposition} we  assume that $\dim\ker\, \si^{\cpl}$ is even or infinite. 


In this representation, we can  define the boundary von Neumann algebra as follows
\beqa
\mathfrak{M}_{\bd, \mrm{KW}}\defeq \{  \e^{\i\phi_{\mrm{KW}}( \p'f) } \,|\, f\in \cX_{\bd} \,\}'',
\eeqa
where $\phi_{\mrm{KW}}$ is defined in (\ref{KW-field}), and it makes sense on $\p'f\in \cX^{\cpl}$. Here the prime denotes the commutant again.
\bet\label{thm:kw} We have $\pi_{\mrm{KW}}(\mfA_{\bulk})''\subset \mathfrak{M}_{\bd, \mrm{KW}}$.
\eet
\proof Due to the relation $(\cX_{\bulk})^{\mrm{cl}}\subset (\Ran\,( \p' \res   \cX_{\bd}))^{\mrm{cl}}$, for any $w\in \cX_{\bulk}$
we can find a sequence of $f_n\in \cX_{\bd}$ s.t.~$\p' f_n\to w$ { in the Hilbert space topology given by $\eta$ and therefore also in the 
topology given by $(\,\cdot\,|\,\cdot\,)_{\mrm{KW}}$ defined in (\ref{KW-scalar-product}). Since
\beqa
\phi_{\mrm{KW}}(x)\defeq \phi_{\mrm{F}}((1+|b|)^{1/2}x\oplus (1-\ov{|b|})^{1/2} \ov{x}), \quad x\in \cX^{\cpl}
\eeqa
and $\|b\|\leq 1$, we obtain from Lemma~\ref{Weyl-operator-approximation} in the appendix that 
$\e^{\i \phi_{\mrm{KW}}( \p'f_n)}\to  \e^{\i\phi_{\mrm{KW}}(w)}$
 in the strong operator topology. This gives $\pi_{\mrm{KW}}(\mfA_{\bulk})\subset \mathfrak{M}_{\bd, \mrm{KW}}$ 
 and thus the claim follows. \qed

\section{Holography on asymptotically $\AdS$ spacetimes}\label{sec:AdS}

\subsection{Notation} If $M$ is a smooth manifold with boundary $\p M$, we denote by $M^\inti$ its interior.  Let $z$ be a smooth boundary-defining function of $\p M$, i.e.~$z>0$ on $M^\inti$, $\p M=\{z=0\}$, and $dz\neq 0$ on $\pM$. Let us recall that given $z$, there exists $W\supseteq\pM$, $\epsilon>0$ and a diffeomorphism $\phi:[0,\epsilon)\times \pM \to W$ such that $z\circ \phi$ agrees with the projection to the first component of $[0,\epsilon)\times \pM$  (see e.g.~\cite[Thm.~9.25]{lee}). We always assume that such $\phi$ is already given and drop it in the notation subsequently. When working close to the boundary we will use various notation proper of $[0,\epsilon) \times \pM$ whenever it is unlikely to cause any confusion. 


On the boundaryless manifold $\p M$ we use the conventional notation $\cD'(\pM)$ for the space of distributions.

The signature of Lorentzian metrics is taken to be $(+,-,\dots,-)$. 

\subsection{Setup} Let us first introduce the setup.  We employ Vasy's definition of asymptotically $\AdS$ spacetimes \cite{vasy}. Let $M$ be an $n$-di\-men\-sio\-nal ($n\geq 2$) smooth manifold $M$ with boundary $\p M$, and suppose $g$ is a Lorentzian metric on its interior  $M^\inti$.

\begin{definition}\label{defads} $(M,g)$ is called an asymptotically anti-de Sitter (a\AdS) spacetime  if near $\p M$, the metric $g$ is of the form
\beq\label{eq:formproduct}
g=\frac{-dz^2+h}{z^2},
\eeq
with $h\in\cf(M;{\rm Sym}^2 T^* M)$ such that with respect to some product decomposition $M=\pM\times [0,\epsilon)_z$ near $\p M$, the restriction $h\traa{\pM}$ is a section of $T^*\pM\otimes T^*\pM$ and is a Lorentzian metric on $\pM$.  
\end{definition}

To account for the fact that null geodesics in $M^\inti$ can hit the boundary, it is useful to introduce the notion of \emph{broken null geodesics} on $M$. They can be defined as the projection to $M$ of the so-called generalized broken bicharacteristics, see \cite{vasy} for the definition and a discussion in the present context.

 Similarly to \cite{vasy}, we make the following two global assumptions:
\begin{enumerate}
\item[$(\rm TF)$] there exists $t\in \cf(M)$ which on each broken null geodesic is either strictly increasing or strictly decreasing, and takes all real values;\vspace{0.1cm}
\item[$(\rm PT)$] topologically, $M=\rr_t\times\Sigma$ for a compact manifold $\Sigma$ with boundary, and $\{t=s\}\times\Sigma$ is spacelike for each $s\in \rr$.
\end{enumerate}

\noindent The basic examples are:

\begin{example} The \emph{universal cover of anti-de Sitter space} can be described as the manifold $M_{\AdS}= \rr_t\times \overline{\mathbb{B}^{n-1}}$ (where $\overline{\mathbb{B}^{n-1}}$ is the closed ball) with metric given by
\[
g_{\AdS}=\frac{ -(1+z^2)^{-1}dz^2 +(1+z^2)dt^2- d\omega^2}{z^2}
\]
on $[0,1)_z \times \rr_t \times \ss_\omega^{n-2}$, and with boundary $\pM_{\AdS}=\{z=0\}$, see e.g.~the discussion in \cite{vasy}. 
\end{example}

\begin{example} The \emph{Poincar\'e patch of anti-de Sitter space} is the manifold $M_{\rm PAdS}=[0,\infty)_z\times \rr^{n-1}$ equipped with the metric
\[
g_{\rm PAdS}=\frac{-dz^2+dt^2-d\bx^2}{z^2}.
\]
\end{example}

We consider the  Klein-Gordon operator on an a$\AdS$ spacetime $(M,g)$:
\beq\label{eq:P}
P\defeq\Box_g + \mass, 
\eeq
where $\Box_g=\frac{1}{\sqrt{|g|}}\p_\mu(\sqrt{|g|} g^{\mu\nu}\p_\nu)$. Throughout the paper we assume that $\nu>0$ (this is the so-called Breitenlohner--Freedman bound \cite{breiten}), so that the analysis in \cite{vasy} applies, cf.~\cite{bachelot,holzegel,warnick1,warnick2,gannot} for related results.

\subsection{Spaces of distributions} We now briefly introduce spaces of distributions which are natural for studying solutions of $Pu=f$ with Dirichlet boundary conditions. We remark that this is less straightforward than on globally hyperbolic spacetimes, since solutions with compactly supported data can reach the boundary $\pM$ (which plays the r\^ole of spatial infinity) at finite times. Thus, special care is needed to capture the correct behaviour at $\pM$.

\medskip
 
We follow \cite{vasy,wrochna}, and refer the reader there for the precise definitions, including a description of how the various spaces are topologized.

Let $L^2(M)=L^2(M,g)$, and let us denote by $(\cdot| \cdot)_{L^2}$ the corresponding inner product.

The space of \emph{extendible distributions} $\cC^{-\infty}(M)$ is defined as the dual of the space of smooth functions vanishing with all derivatives at $\pM$.

The \emph{zero-Sobolev space} of order $1$ is defined by
\[
 H^1_0(M)=\big\{  u\in \cC^{-\infty}(M)\, | \, Qu\in L^2(M)\ \forall Q\in \Diff^1_0(M)  \big\},  
\]
where $\Diff^1_0(M)$ is the space of all vector fields vanishing at $\pM$. Furthermore, one sets $H^0_0(M)=L^2(M)$, and $H^{-1}_0(M)$ is the dual of $H^{1}_0(M)$, defined using the $L^2(M)$ pairing.

For $k\in\{-1,0,1\}$, $H^{k,\infty}_{0,\b}(M)$ is the  Fr\'echet space of \emph{conormal distributions} respective to $H^k_0(M)$, i.e.,
\[
 H^{k,\infty}_{0,\b}(M)=\big\{  u\in H_0^k(M)\, | \, Qu\in H_0^k(M)\ \forall Q\in\Diff_\b(M)  \big\}, 
\]
where $\Diff_\b(M)$ consists of differential operators which are polynomials of vector fields \emph{tangent to $\pM$}. The space $H^{-k,-\infty}_{0,\b}(M)$ is then defined as the dual of $H^{k,\infty}_{0,\b}(M)$.

We will also consider the subspace $H^{k,\infty}_{0,\b,\c}(M)$ of $H^{k,\infty}_{0,\b}(M)$ (resp.~the subspace $H^{k,-\infty}_{0,\b,\c}(M)$ of $H^{k,-\infty}_{0,\b}(M)$) that consists of compactly supported elements. Here, `compact support' is meant in the sense of the manifold with boundary $M$, in particular overlapping $\pM$ is allowed. Since in our setup, spatial directions are compactified, `compact support' is interpreted as `supported in a finite interval of time'.

Furthermore, let $H^{k,\infty}_{0,\b,\loc}(M)$ be the space of distributions such that their localization with test functions (smoothly extendible across $\pM$) belongs to $H^{k,\infty}_{0,\b}(M)$. We define $H^{k,\infty}_{0,\b,\c}(M)$ similarly. Then, once appropriately topologized as inductive limits of  Fr\'echet spaces, $H^{k,\infty}_{0,\b,\loc}(M)$ is in duality with $H^{-k,-\infty}_{0,\b,\c}(M)$, and $H^{k,\infty}_{0,\b,\c}(M)$ is in duality with $H^{-k,-\infty}_{0,\b,\loc}(M)$.

Let us remark that over the interior $M^\inti$, $H^{k,\infty}_{0,\b,\loc}(M)$ coincides with $\cf(M^\inti)$ (the usual space of smooth functions on the interior) and $H^{k,-\infty}_{0,\b,\loc}(M)$ with $\cD'(M^\inti)$. For our purpose, the spaces $H^{k,\infty}_{0,\b,\c}(M)$ provide a good replacement for the space of test functions $\cf_\c(M^\inti)$ used in the globally hyperbolic case.

\subsection{Symplectic space of solutions} Let us denote by $\Hpm^{k,\infty}(M)$ the space of future/past supported elements of $\Hl^{k,\infty}(M)$, i.e.
\beq\label{eq:dsdfsd}
\Hpm^{k,\infty}(M)= \big\{ u\in \Hl^{k,\infty}(M)\, | \,  \supp u \subset \{\pm t\geq \pm t_0\} \mbox{ for some } t_0\in\rr\big\}.  
\eeq
By {\cite[Thm.~1.6]{vasy}}, there exist \emph{Dirichlet retarded/advanced propagators}, denoted by respectively $P_\pm^{-1}$,  i.e.~unique continuous operators
\beq\label{pp0}
P_\pm^{-1}: \Hpm^{-1,\infty}(M)\to \Hpm^{1,\infty}(M)
\eeq
that satisfy $P P_\pm^{-1} = \one$ on $\Hpm^{-1,\infty}(M)$ and $P_\pm^{-1} P = \one$ on $\Hpm^{1,\infty}(M)$. 

The difference of the two propagators,
\beq\label{eq:defG}
G\defeq P_+^{-1}-P_-^{-1} : \Hc^{-1,\infty}(M)\to \Hl^{1,\infty}(M),
\eeq
is an analogue (in general not unique, since different boundary conditions are possible) of the \emph{Pauli-Jordan propagator} (also called \emph{causal propagator} or \emph{commutator function}) on globally hyperbolic spacetimes.

It is straightforward to show the following proposition, which is an analogue of \cite[Prop.~3.1]{wrochna} for real-valued function spaces. 

\begin{proposition}\label{prop:symp} The $\rr$-bilinear form $(\cdot|  G\cdot)_{L^2}$ induces a non-degenerate symplectic form on the quotient space
\beq\label{eq:iso1}
\cX\defeq\frac{ \Hc^{-1,\infty}(M;\rr) }{P  \Hc^{1,\infty}(M;\rr)}.
\eeq
\end{proposition}
 As it is the inductive limit of  Fr\'echet spaces, $\Hc^{-1,\infty}(M;\rr)$ is barreled. By \cite[Prop.~33.1]{treves}, $\cX$ equipped with the quotient topology is a  barreled space.

We denote by $\sigma$ the symplectic form from Proposition \ref{prop:symp}. The symplectic space $(\cX,\sigma)$ represents the classical field theory in the bulk: it can indeed be shown that $(\cX,\sigma)$ is isomorphic to a suitable symplectic space of solutions of $Pu=0$ \cite{wrochna} (cf.~\cite{DF17} for an approach specialized to the Poincar\'e patch of AdS, though allowing for different boundary conditions).

\subsection{Boundary data of classical fields}\label{ss:holo} Let $\nu_\pm=\frac{n-1}{2}\pm \nu$ be the two \emph{indicial roots} of $P$. The next proposition is a variant of \cite[Prop.~8.10]{vasy} shown in \cite[Prop.~3.7]{wrochna}.

\begin{proposition}\label{mimi} Suppose that $u\in H^{1,-\infty}_{0,\loc}(M)$ solves $Pu=0$. Then $u$ is of the form
\beq\label{formu}
\bea
u=z^{\nu_+} v, \ \ &v\in \cf\dv.
\eea
\eeq
Furthermore, the map $u\mapsto v\traa{\pM}$ is continuous in the $H^{1,-\infty}_{0,\loc}(M)$ and $\cD'(\pM)$ topologies. 
\end{proposition}

In the present setting, the \emph{classical bulk-to-boundary} map is the continuous map   
\beq
\bea
\p_+  : z^{\nu_+}\cf\dv & \to \cD'(\pM) \\
\p_+ u & \defeq (z^{-\nu_+} u)\traa{\pM}.
\eea
\eeq
On solutions of $Pu=0$, $\p_+$ coincides with the map $u\mapsto v\traa{\pM}$ from Proposition \ref{mimi}. By identifying elements of $\cX'$ (the dual of the quotient space $\cX$ defined in \eqref{eq:iso1}) with solutions of $Pu=0$, we also obtain a map $\p_+:\cX'\to\cD'(\pX)$, which is continuous by Proposition \ref{mimi}. An important question for us is whether it is injective, at least in some weak sense.

\begin{definition}\label{def:cont} For $O\subset \pM$ open we define its \emph{domain of uniqueness} $V(O)\subset M$ to be the maximal open set such that for any $u\in H^{1,-\infty}_{0,\b,\loc}(M)$ solving $Pu=0$ on $M$ we have the implication
\beq
\big(  \partial_+ u=0 \mbox{ on } O \big) \Rightarrow  \big( u =0 \mbox{ on } V(O)\big).
\eeq 
We say that the \emph{unique continuation property} holds true on $O$ if $V(O)\neq \emptyset$.
\end{definition}

We now make the connection with the abstract framework introduced in Section \ref{sec:abstract}. For $O\subset\pM$ open and $V\subset M$ open we set
\beqa
\quad\quad\quad \cX_{\bd}(O)\defeq\cf_{\rm c}(O), \ \ \cX_{\bulk}(V)\defeq \frac{\Hc^{-1,\infty}(V;\rr)}{P  \Hc^{1,\infty}(M;\rr)\cap
 \Hc^{-1,\infty}(V; \rr )}. \label{quotient}
\eeqa
In particular, $\cX_\bulk(M)=\cX$, where $\cX$ was defined in \eqref{eq:iso1}. We remark here that given a quotient space $E/F$ and a subspace $K\subset E$, we have a canonical injection $(E\cap K)/(F\cap K)\to E/F$ given by $f+(F\cap K)\mapsto f+F$. In this sense we can treat $\cX_{\bulk}(V)$ as a subspace of $\cX_{\bulk}$.

Let $\eta\in L_{\rm s}(\cX,\cX')$ be continuous and suppose it satisfies the positivity condition
\beq
\eta\geq 0 \textrm{ and } |v_1\cdot \sigma v_2|\leq (v_1\cdot \eta v_1)^{1/2} (v_2\cdot \eta v_2)^{1/2}, \quad v_1,v_2\in \cX. \label{positivity-property2}
\eeq
 Note that since $\cX$ is a quotient space, well-definiteness of $\eta$ means that $\eta$ is a bi-solution, i.e.
\beq\label{eq:etac1}
 P\circ\eta=0 \mbox{ on } \Hc^{-1,\infty}(M) \mbox{ and } \eta\circ P=0 \mbox{ on } \Hc^{1,\infty}(M).
\eeq
Continuity means that
\beq\label{eq:etac2}
\eta:\Hc^{-1,\infty}(M)\to\Hl^{1,-\infty}(M) \mbox{ is continuous}.
\eeq
The properties \eqref{positivity-property2}--\eqref{eq:etac2} are satisfied e.g.~by the covariances of any \emph{holographic Hadamard state}, see \cite{wrochna}. In particular this is true for the covariances of ground states on static aAdS spacetimes.

We define as in Subsect.~\ref{ss:abstract} the $C^*$-algebras
\beq\bea\label{eq:nets}
\mfa_{\bd}(O)&\defeq\mrm{CCR} \big((\Ran\,( \p' \res   \cX_{\bd}))^{\mrm{cl}}, \si^{\cpl}\big), \\
\mfa_{\bulk}(V)&\defeq\mrm{CCR} \big(\cX^{\mrm{cl}}_{\bulk}, \si^{\cpl}\big),
\eea\eeq
where now $\cX_\bd=\cX_\bd(O)$, $\cX_\bulk=\cX_\bulk(V)$, and $\p=\p_+$. The completions and closures refer to the topology induced by $\eta$. Note that using the completion is necessary because distributions of the form $\p' f$ have a `Dirac delta singularity' at the boundary, and are therefore too singular to belong to $\cX$.  

We remark that \eqref{eq:nets} defines isotonic nets of algebras, i.e., $O_1\subset O_2$ implies $\mfa_{\bd}(O_1)\subset \mfa_{\bd}(O_2)$ and $V_1\subset V_2$ implies $\mfa_{\bulk}(V_1)\subset \mfa_{\bulk}(V_2)$.

By applying Theorem \ref{Main-result} we obtain:

\begin{theorem}\label{maintheorem} Let $O\subset \pM$ be an open set, and let $V(O)$ be  its domain of uniqueness (see Definition \ref{def:cont}). Let $\mfa_{\bd}(O)$ and $\mfa_{\bulk}(V(O))$ be defined by \eqref{eq:nets} for some $\eta$  satisfying \eqref{positivity-property2}--\eqref{eq:etac2}. Then
\beq
\mfa_{\bd}(O)\supset\mfa_{\bulk}(V(O)). 
\eeq
\end{theorem} 

This result can be reformulated in a Kay-Wald representation by analogy with Theorem \ref{thm:kw}.

\medskip

  We point out,  that in the case when $(M,g)$ is analytic,  our unique continuation property $V(O)\neq \emptyset$, where $V(O)$ is some bulk neighborhood of $O$, holds true by Holmgren's theorem\footnote{See e.g.~\cite[Thm.~8.6.5]{hoermander}.}, for $O$ arbitrarily small. Thus, our Theorem \ref{maintheorem} is directly applicable  in the analytic case.

It is an interesting question for future research  if our result also applies to spacetimes which are  smooth, but not analytic.  
We restrict ourselves here to several remarks in favour of such conjecture.

 A weaker variant of the unique continuation property was established recently by Holzegel and Shao \cite{HS1,HS2}, for more restrictive classes of solutions of $Pu=0$. For convenience our terminology refers to Dirichlet boundary conditions and linear fields exclusively. In this narrowed setting, the results of Holzegel and Shao can be summarized as follows. 
\begin{enumerate}
\item On a large class of a$\AdS$ spacetimes, the local unique continuation property for sufficiently smooth solutions $u$ holds true on $O$ if  $[-T,T]\times \p\Sigma\subset O$ for $T$ sufficiently large (enough for null geodesics to refocus in that interval of time). As of now, there is no known general estimate on the size of $V(O)$.  (See condition (PT) above for the definition of $\Sigma$).
\item In the exact AdS case, $V(O)=M$ for $O$ as above.
\end{enumerate}

\nin  We emphasize that the statement and the proof of (1) do not use analyticity and admits generalizations to non-linear classical fields.  

It is expected that the statement (2) should hold true for some larger class of spacetimes. On the other hand, the assumption on the minimal size of $O$ in (1) is expected to be necessary if one does not assume analyticity. 

 To apply our main result one needs to know if for $O$ as in Holzegel and Shao's work, the unique continuation property holds true for all $u\in H^{1,-\infty}_{0,\b,\loc}(M)$, rather than merely for all $u$ smooth. In view of Remark \ref{theremark} it would actually be sufficient to prove this for all $u\in H^{1,-\12}_{0,\b,\loc}(M)$ (see \cite{vasy} for the definition of that space), under the assumption that $\eta$ maps continuously $H^{-1,\12}_{0,\b,\c}(M)\to H^{1,-\12}_{0,\b,\loc}(M)$ (which is expected to be the case in applications). Outside of the context of anti-de Sitter spacetimes it would be also interesting to know if the results of Ionescu and Klainerman \cite{klainerman} and Lerner \cite{lerner} on unique continuation across characteristic surfaces can be extended to some low regularity classes of solutions. 

\appendix
\section{Quasi-free states and Kay-Wald representation}\label{theappendix}
\nin We follow here the book \cite{derger} and the lectures notes \cite{gerardnotes}.

\subsection{Fock spaces}
Let $\mfh$ be a complex Hilbert space and $\Ga(\mfh)$ the symmetric Fock space, with the vacuum vector denoted by $\Om_{\mrm{vac}}$. 
We denote by $a^{(*)}(h)$, $h\in \mfh$, the
creation and annihilation operators. They satisfy
\beqa
[a(h_1), a^*(h_2)]=(h_1|h_2)\one, \quad h_1, h_2\in \mfh,
\eeqa
and all the other commutators vanish. We also define 
\beqa
\phi_{\mrm{F}}(h)\defeq \frac{1}{\sqrt{2}} \big(a^*(h)+a(h)\big) \textrm{ and } W_{\mrm{F}}(h)\defeq \e^{\i \phi_{\mrm{F}}(h)}.
\eeqa
From this we compute
\beq\bea
\, [ \phi_{\mrm{F}}(h_1), \phi_{\mrm{F}}(h_2)  ]=\i \mrm{Im}(h_1|h_2)\one, \\  W_{\mrm{F}}(h_1)  W_{\mrm{F}}(h_2)=\e^{-\frac{\i}{2} \mrm{Im}(h_1|h_2)  } 
W_{\mrm{F}}(h_1+h_2).
  \eea\eeq
The next lemma is standard, see e.g.~\cite[Thm.~X.41~(d)]{RS2}.
\bel\label{Weyl-operator-approximation} Suppose that $h_n\to h$ in the norm topology of $\mfh$. Then, for any $\Psi\in \Ga(\mfh)$, $W_{\mrm{F}}(h_n)\Psi\to W_{\mrm{F}}(h)\Psi$.
\eel




\subsection{CCR Algebras}

A bilinear form $\si\in L(\cX, \cX')$ is called pre-symplectic if it is antisymmetric. 
If, in addition, it is non-degenerate, then it is called symplectic. 

Let $(\cX,\si)$  be a (pre)-symplectic space.
The \emph{polynomial CCR algebra} is the $*$-algebra  generated by abstract scalar fields $\phi(v)$, $v \in \cX$, satisfying the relations
\beqa
& &\phi(v_1+\la v_2)=\phi(v_1)+\la\phi(v_2), \\
& &\phi^*(v)=\phi(v),\\
& &[\phi(v_1), \phi(v_2)]=\i (v_1\cdot\si v_2) \one.
\eeqa 
Now, the \emph{Weyl CCR algebra} $\mrm{CCR}(\cX, \si )$ is the $C^*$-algebra generated by
$\cX \ni v\mapsto W(v)$ satisfying
\beq\bea
W(v_1)W(v_2)=\e^{-\frac{\i}{2} v_1\cdot \si v_2} W(v_1+v_2), \\ W(v)^*=W(-v), \quad W(0)=\one.
\eea\eeq

\subsection{Quasi-free states and their GNS representations}

A quasi-free state $\om$ on $\mfa=\mrm{CCR}(\cX, \si)$ has the property
\beqa
v\mapsto \om(W(v))=\e^{-\12 v\cdot \eta v},
\eeqa
where $\eta: \cX \to \cX'$ is the covariance. Positivity of the state is equivalent to
\beq
\eta\geq 0 \textrm{ and } |v_1\cdot \sigma v_2|\leq (v_1\cdot \eta v_1)^{1/2} (v_2\cdot \eta v_2)^{1/2}, \quad v_1,v_2\in \cX. \label{positivity-property}
\eeq
If $\si$ is symplectic, we have  $\ker\, \eta=\{0\}$ by (\ref{positivity-property}).  We denote by $\cX^{\cpl}$ the
completion of $\cX$ for $(v\cdot \eta v)^{1/2}$ which is a real Hilbert space. The extension $\si^{\cpl}$ of the
symplectic form $\si$ is bounded on $\cX^{\cpl}$ but may become degenerate. We also recall that $\om$
induces a unique quasi-free state $\om^{\cpl}$ on $\mrm{CCR}(\cX^{\cpl}, \si^{\cpl})$ and we denote the resulting  covariance by $\eta^{\cpl}$.

In order to construct a complex Hilbert space out of the real Hilbert space $\cX^{\cpl}$, $\si^{\cpl}$ and $\eta^{\cpl}$ we need the following fact:
\bep Let $\eta^{\cpl} $ be the real covariance of a quasi-free state on the algebra  $\mrm{CCR}(\cX^{\cpl}, \si^{\cpl})$ such that $\eta^{\cpl}$ is non-degenerate and $\cX^{\cpl}$
is complete for the norm $(v\cdot \eta v)^{1/2}$. Then if $\dim\ker\, \si^{\cpl}$ is even or infinite, there exists 
an anti-involution $\mrm{j}$ on $\cX^{\cpl}$ s.t. $(\eta^{\cpl}, \mrm{j})$ is K\"ahler, that is, $\eta^{\cpl}$ is positive and $-\eta^{\cpl} \mrm{j}$ is symplectic.
\eep
The first step of the proof of this proposition is to conclude from (\ref{positivity-property})
that $\si^{\cpl}=2\eta^{\cpl} b$, where $b\in L(\cX^{\cpl})$ is an antisymmetric operator s.t.~$\|b\|\leq 1$.
Then, one uses the polar decomposition of $b$ on the orthogonal complement of $\Ker b$ and an arbitrary anti-involution on $\Ker b$  to construct $\mrm{j}$.

We equip $\cX^{\cpl}$ with the complex structure $\mrm{j}$ as above and define the scalar product
\beqa
(v_1|v_2)_{\mrm{KW}}\defeq v_1\cdot \eta^{\cpl} v_2-\i v_1\cdot \eta^{\cpl} \mrm{j} v_2, \label{KW-scalar-product}
\eeqa
which turns $(\cX^{\cpl}, \mrm{j}, (\, \cdot\, |\, \cdot\, )_{\mrm{KW}})$ into a complex Hilbert space which we will 
denote by $\cX_{\mrm{KW}}$. We denote by $\ov{\cX}^{\cpl}_{\mrm{KW}}$ the complex conjugate space, which is obtained by replacing $\mrm{j}$ with $-\mrm{j}$
in the above construction. $\cX_{\mrm{KW}}^{\cpl}$ and $\ov{\cX}^{\cpl}_{\mrm{KW}}$ coincide as real vector spaces and the resulting (antilinear) identity map is denoted
by $x\to \ov{x}$. Also given a linear map  $a: \cX_{\mrm{KW}}^{\cpl} \to \cX_{\mrm{KW}}^{\cpl}$ we define $\ov{a}: \ov{\cX}^{\cpl}_{\mrm{KW}} \to \ov{\cX}^{\cpl}_{\mrm{KW}}$ by $\ov{a}\, \ov{x}=\ov{ax}$. Using these definitions, we set
\begin{align}
\mfh_{\mrm{KW}}&\defeq \cX_{\mrm{KW}}\oplus 1_{\real\backslash 1}(\ov{|b|})\ov{\cX_{\mrm{KW}}}, \\
\phi_{\mrm{KW}}(v)&\defeq \phi_{\mrm{F}}((1+|b|)^{1/2}v\oplus (1-\ov{|b|})^{1/2} \ov{v}), \quad v\in \cX^{\cpl},        \label{KW-field}
\end{align}
acting on $\Ga(\mfh_{\mrm{KW}})$. For the Weyl algebra one then obtains as in \cite[Prop.~4.18]{gerardnotes} the \emph{Kay-Wald representation}:
 \bep\label{GNS-proposition} The triple $(\hil_{\mrm{KW}}, \pi_{\mrm{KW}}, \Om_{\mrm{KW}})$ defined by
 \[
 \hil_{\mrm{KW}}\defeq \Ga(\mfh_{\mrm{KW}}), \quad \pi_{\mrm{KW}}(W(v))=\e^{\i\phi_{\mrm{KW}}(v) }, \ \ v\in \cX^{\cpl},  \quad \Om_{\mrm{KW}}=\Om_{\mrm{vac}}  
 \]
 is the GNS triple of the quasi-free state $\om$ on $\mrm{CCR}(\cX^{\cpl}, \si^{\cpl})$. 
 \eep




\end{document}